\documentstyle[12pt]{article}

\catcode`\@=11
\long\def\@makefntext#1{
\protect\noindent \hbox to 3.2pt {\hskip-.9pt
$^{{\ninerm\@thefnmark}}$\hfil}#1\hfill}		%CAN BE USED

\def\@makefnmark{\hbox to 0pt{$^{\@thefnmark}$\hss}}  %ORIGINAL

\def\ps@myheadings{\let\@mkboth\@gobbletwo
\def\@oddhead{\hbox{}
\rightmark\hfil\ninerm\thepage}
\def\@oddfoot{}\def\@evenhead{\ninerm\thepage\hfil
\leftmark\hbox{}}\def\@evenfoot{}
\def\sectionmark##1{}\def\subsectionmark##1{}}

\renewcommand{\thefootnote}{\fnsymbol{footnote}}

\newcounter{sectionc}\newcounter{subsectionc}\newcounter{subsubsectionc}
\renewcommand{\section}[1] {\vspace*{0.6cm}\addtocounter{sectionc}{1}
\setcounter{subsectionc}{0}\setcounter{subsubsectionc}{0}\noindent
	{\normalsize\bf\thesectionc. #1}\par\vspace*{0.4cm}}
\renewcommand{\subsection}[1] {\vspace*{0.6cm}\addtocounter{subsectionc}{1}
	\setcounter{subsubsectionc}{0}\noindent
	{\normalsize\it\thesectionc.\thesubsectionc. #1}\par\vspace*{0.4cm}}
\renewcommand{\subsubsection}[1]
{\vspace*{0.6cm}\addtocounter{subsubsectionc}{1}
\noindent {\normalsize\rm\thesectionc.\thesubsectionc.\thesubsubsectionc.
	#1}\par\vspace*{0.4cm}}

\newcounter{appendixc}
\newcounter{subappendixc}[appendixc]
\newcounter{subsubappendixc}[subappendixc]

\renewcommand{\appendix}[1] {\vspace*{0.6cm}
        \refstepcounter{appendixc}
        \setcounter{figure}{0}
        \setcounter{table}{0}
        \setcounter{equation}{0}
        \renewcommand{\thefigure}{\Alph{appendixc}.\arabic{figure}}
        \renewcommand{\thetable}{\Alph{appendixc}.\arabic{table}}
        \renewcommand{\theappendixc}{\Alph{appendixc}}
        \renewcommand{\theequation}{\Alph{appendixc}.\arabic{equation}}
        \noindent{\bf Appendix \theappendixc #1}\par\vspace*{0.4cm}}

\def\abstracts#1{{
\centering{\begin{minipage}{12.2truecm}\footnotesize\baselineskip=12pt\noindent
	\centerline{\footnotesize ABSTRACT}\vspace*{0.3cm}
	\parindent=0pt #1
	\end{minipage}}\par}}

\renewenvironment{thebibliography}[1]
	{\begin{list}{\arabic{enumi}.}
	{\usecounter{enumi}\setlength{\parsep}{0pt}
\setlength{\leftmargin 1.25cm}{\rightmargin 0pt}
	 \setlength{\itemsep}{0pt} \settowidth
	{\labelwidth}{#1.}\sloppy}}{\end{list}}

\newcounter{itemlistc}
\newcounter{romanlistc}
\newcounter{alphlistc}
\newcounter{arabiclistc}

\newcommand{\fcaption}[1]{
        \refstepcounter{figure}
        \setbox\@tempboxa = \hbox{\footnotesize Fig.~\thefigure. #1}
        \ifdim \wd\@tempboxa > 6in
           {\begin{center}
        \parbox{6in}{\footnotesize\baselineskip=12pt Fig.~\thefigure. #1}
            \end{center}}
        \else
             {\begin{center}
             {\footnotesize Fig.~\thefigure. #1}
              \end{center}}
        \fi}

\newcommand{\tcaption}[1]{
        \refstepcounter{table}
        \setbox\@tempboxa = \hbox{\footnotesize Table~\thetable. #1}
        \ifdim \wd\@tempboxa > 6in
           {\begin{center}
        \parbox{6in}{\footnotesize\baselineskip=12pt Table~\thetable. #1}
            \end{center}}
        \else
             {\begin{center}
             {\footnotesize Table~\thetable. #1}
              \end{center}}
        \fi}

\def\@citex[#1]#2{\if@filesw\immediate\write\@auxout
	{\string\citation{#2}}\fi
\def\@citea{}\@cite{\@for\@citeb:=#2\do
	{\@citea\def\@citea{,}\@ifundefined
	{b@\@citeb}{{\bf ?}\@warning
	{Citation `\@citeb' on page \thepage \space undefined}}
	{\csname b@\@citeb\endcsname}}}{#1}}

\newif\if@cghi
\def\cite{\@cghitrue\@ifnextchar [{\@tempswatrue
	\@citex}{\@tempswafalse\@citex[]}}
\def\citelow{\@cghifalse\@ifnextchar [{\@tempswatrue
	\@citex}{\@tempswafalse\@citex[]}}
\def\@cite#1#2{{$\null^{#1}$\if@tempswa\typeout
	{IJCGA warning: optional citation argument
	ignored: `#2'} \fi}}

 1
 1
 1

\font\ninerm=cmr9

\begin{document}

\newcommand{\st}{\scriptstyle}
\newcommand{\sst}{\scriptscriptstyle}
\newcommand{\mco}{\multicolumn}
\newcommand{\epp}{\epsilon^{\prime}}
\newcommand{\vep}{\varepsilon}
\newcommand{\ra}{\rightarrow}
\newcommand{\ppg}{\pi^+\pi^-\gamma}
\newcommand{\vp}{{\bf p}}
\newcommand{\ko}{K^0}
\newcommand{\kb}{\bar{K^0}}
\newcommand{\al}{\alpha}
\newcommand{\ab}{\bar{\alpha}}
\def\be{\begin{equation}}
\def\ee{\end{equation}}
\def\bea{\begin{eqnarray}}
\def\eea{\end{eqnarray}}
\def\CPbar{\hbox{{\rm CP}\hskip-1.80em{/}}}%temp replacement due to no font

%\centerline{\normalsize\bf WORLD SCIENTIFIC PUBLISHING COMPANY}
%\baselineskip=22pt
January, 1995 \hfill LBL-36740\\
\vskip 0.2in
\centerline{\normalsize\bf DISORIENTED CHIRAL CONDENSATE\footnote{This
work was supported
by the Director, Office of Energy
Research, Office of High Energy and Nuclear Physics,
Divisions of High
Energy Physics
 of the U.S. Department of Energy under Contract
DE-AC03-76SF00098
 and by the Natural Sciences and
Engineering Research Council of Canada.}}
%\baselineskip=16pt
%\centerline{\normalsize\bf MANUSCRIPT BY COMPUTER}
%\centerline{\footnotesize\sf (For subsequent 20\% photoreduction to 17.8
%$\times$ 11.9 cm text area)
%\footnote{The \LaTeX\ source file for this document may
%be used as a template for your article, and can be requested by e-mailing
%{\sf worldscp@singnet.com.sg}.}}
%\vfill
\vspace*{0.6cm}
\centerline{\footnotesize ZHENG HUANG\footnote{Talk presented
at the Beyond The Standard Model IV, 13-18 December 1994, Lake Tahoe,
California}}
\baselineskip=13pt
\centerline{\footnotesize\it Theoretical Physics Group,
Lawrence Berkeley Laboratory
}
\baselineskip=12pt
\centerline{\footnotesize\it University of California,
Berkeley, CA 94720, USA}
\centerline{\footnotesize E-mail: huang@theorm.lbl.gov}
%\vspace*{0.3cm}
%\centerline{\footnotesize and}
%\vspace*{0.3cm}
%\centerline{\footnotesize SECOND AUTHOR'S NAME}
%\baselineskip=13pt
%\centerline{\footnotesize\it Group, Company, Address, City, State ZIP/Zone,
%Country}
%\vfill
\vspace{0.4cm}
\abstracts{
The current theoretical understanding of disoriented chiral
condensate is briefly
reviewed. I discuss the basic idea, the formation mechanism and experimental
signatures of DCC in high energy collisions.
}

%\vspace*{0.6cm}
\normalsize\baselineskip=15pt
\setcounter{footnote}{0}
\renewcommand{\thefootnote}{\alph{footnote}}
\section{Basic Idea}
The spontaneous symmetry-breaking mechanism plays a very important role
in high energy physics. It is known that there are
at least two occurrences of such phenomenon at work in the standard model:
the electroweak symmetry-breaking and the chiral symmetry-breaking,
in which the observed asymmetries are attributed entirely to the
vacuum states of our universe. But how do we test this idea directly?
Is there any way that we can create a suitable condition under which the
vacuum state is disturbed for a small region of space-time so that
we may be able to observe some quite
different excitations and domain structures in the vacuum?
%This may be
%possible for the chiral symmetry-breaking in strong interactions given the
%present high energy physics facilities.

Let us examine the possibility for the chiral symmetry in strong interactions.
Suppose a very high energy proton-proton or nucleus-nucleus collision in cosmic
rays or in Tevatron or RHIC collider. Occasionally, the collision creates a
large number of low energy (small $p_t$) particles, mainly pions.
These strongly interacting particles initially
populate in a small phase space and rescatter many times before
leaving the system, heating up a small local space-time region $\Omega$.
The initial fluctuations may or may not reach a thermal equilibrium.
Outside $\Omega$, it is the normal vacuum state.
The pressure difference between
the interior and the exterior caused by the initial fluctuations results in a
rapid expansion, especially in the beam direction. The evolution of the
fluctuations involves a rolling down from the top of potential to the
``valley''
of the potential, governed mainly by the classical equation of motion.
Such a ``rolling-down'' may be unstable for long wavelength modes, which
will be exponentially amplified.  The
domains of different vacuum structure may emerge in the interior, leading to
a quite different characteristic of the pion productions when these domains
decay. The formation of ``disoriented'' domains is referred to as the
``disoriented chiral condensate (DCC)'' \cite{dcc}.
%In essence it bears
%some analogy to the chaotic inflation scenario in the early universe.

\section{Formation Mechanism}
\subsection{The Model}
How do we tackle the problem at hand since the system contains strongly
interacting particles and the standard perturbation theory fails? Fortunately,
we are only dealing with low energy hadrons, there is a very powerful effective
theory of QCD at low energy, that is, the $\sigma$-model. There are two
versions of the $\sigma$-model: the linear model and the non-linear
model. They are shown to be equivalent to each other at low energy, i.e.\ both
of them satisfy the low energy theorem \cite{low}, which is uniquely
 determined by the symmetry (O(4)). We would like to also include
the oscillation effects between the $\sigma$ field and pion fields.
Therefore, the
linear $\sigma$-model is more appropriate as long as we confine ourself
to the low energy modes. The scalar potential looks like
a Higgs potential in the electroweak standard model except for a very large
self-coupling $\lambda$. Remember, however, that a perturbative series
corresponds to a momentum expansion, which is well justified when the energy
is small compared to ${4\pi f_\pi}\sim 1$ GeV. For example, the partial wave
amplitudes for $\pi \pi$ scatterings, given by the low energy theorem, are
\begin{eqnarray}
a_{00}=\frac{s}{16\pi^2f_\pi^2}\quad ,\quad a_{11}=\frac{s}{96\pi^2f_\pi^2}
\quad ,\quad a_{20}=-\frac{s}{32\pi^2f_\pi^2}
\end{eqnarray}
which are clearly $\lambda$-independent.

Therefore, as long as the energy of $\sigma$ and $\pi$'s is small,
we can ignore the
quantum loop effects and treat the system as classical (
However, it is argued that the quantum effects can be important in the early
evolution of the system before $\sigma$ and $\pi$'s decouple from each other
\cite{cooper}).
As the system expands,
the typical energy-momentum becomes even smaller, the evolution of the system
is
thus governed by the classical equations of motion. However, we should point
out that an initial low energy configuration is our starting point. We have not
started from scratch and calculated the cross section, e.g.,
$\sigma (pp\rightarrow 10^3 \pi$'s, $p_t< 200$ MeV) which necessarily involves
quantum processes. It is only the fact that most particles created in
a given event are dominantly soft pions in hadron-hadron collisions. In a real
experiment, the condition of a low energy is guaranteed by a physical $p_t$
cut from above ($p_t<200$ MeV). In most of theoretical simulations, such
condition is automatically met by choosing a lattice spacing $a\sim 1$ fm.

\subsection{Equations of Motion}
We now have a well-defined problem: given an initial condition, solving the
classical equations of motion.
In the standard linear $\sigma$-model, the equations of
motion are given by,
\begin{equation}
%%%\left ( \frac{\partial ^2}{\partial t^2} - \nabla ^2 \right ) \phi
\Box \phi
= \lambda(f_\pi^2 - \phi^2 )\phi  +H n_{\sigma},\label{em}
\end{equation}
where $\phi \equiv (\sigma, \mbox{\boldmath $\pi$})$ is a vector
in internal space, $n_{\sigma}=(1,\mbox{\bf 0})$, and
$Hn_{\sigma}$ is an explicit chiral symmetry breaking term
due to finite quark masses.
%We shall use $\lambda=19.97$, $f_\pi=87.4$ MeV,
%and $H= (119~{\rm MeV})^3$, which implies $m_\pi = 135$ MeV
%and $m_\sigma = 600$ MeV.
Our goal is to solve Eq. (\ref{em}).

What kind of initial conditions that one may choose? The most general ones
are random gaussian distributions
allowed by a typical energy density or temperature
(just like in the chaotic inflation in the early
universe), where the scalar fields take on different
values in different regions in $\Omega$. To specify a gaussian form, one has to
choose a mean value $\langle \phi\rangle$ over the spatial
volume of $\Omega$ and
a magnitude of the fluctuations $\delta \phi^2$. Let us separate $\phi$
into two parts
\begin{eqnarray}
\phi =\langle \phi \rangle +\delta\phi ,
\end{eqnarray}
where by definition $\langle \delta\phi \rangle =0$. The averaged scalar
potential is
\begin{eqnarray}
\langle V\rangle =\frac{\lambda}{4}
(\langle\phi\rangle^2  + \langle\delta\phi^2\rangle -f_\pi^2 )^2
-H\langle\sigma\rangle .
\end{eqnarray}
If  $\langle\delta\phi^2\rangle$ is to be replaced by a thermal fluctuation
where $\langle\delta\phi^2\rangle =T^2/4$, one recovers the well-known one
loop effective potential at finite temperature.
The instability occurs whenever
$\langle\phi\rangle^2  + \langle\delta\phi^2\rangle -f_\pi^2 <0$.

\subsection{Initial Fluctuations}
The existence of $\langle\delta\phi^2\rangle$ renders the effective potential
for the mean field $\langle\phi\rangle$ different from a zero temperature one.
$\langle\delta\phi^2\rangle -f_\pi^2$ determines the temporal shape of the
potential. There are two scenarios to generate an unstable
``rolling-down''. In an
``annealing'' scenario \cite{anneal}, the system is assumed to be in a
thermal equilibrium, at least initially. The initial value of
$\langle\phi\rangle$ has to match with the equilibrium position (the minimum)
 of the potential
determined by initial value of $\langle\delta\phi^2\rangle$. For example, if
initially $\langle\delta\phi^2\rangle -f_\pi^2>0$, the chiral symmetry is
restored, then the initial value $\langle\phi\rangle \simeq 0$ (approximately
due to the explicit symmetry-breaking). The rolling-down is generated as the
system expands, the $\langle\delta\phi^2\rangle$ decreases and thus the
equilibrium position changes. Such a situation is best realized in a first
order phase transition when the minimum changes discontinuously and when the
potential is sufficiently flat.
In the problem we have here, it seems difficult to achieve a sufficient
rolling-down time in the annealing case.

The second scenario is the ``quenching'' mechanism \cite{quench} which does not
have much to do with the phase transition. There is no thermal equilibrium at
the initial stage, the mean field $\langle\phi\rangle$ is displaced from the
minimum of the potential and rolls down
to the valley. As long as $|\langle\phi\rangle|$ is smaller than
$|\langle\phi_{\rm eq}\rangle|$ at which the effective potential is minimized,
 the instability develops and the long wavelength
modes are exponentially amplified. The expansion causes the fluctuations to
decrease, which tends to increase the roll-down time.
It is clear that such a situation will not be possible if the fluctuations
$\langle\delta\phi^2\rangle > f_\pi^2$ which renders the chiral symmetry
restore and $|\langle\phi_{\rm eq}\rangle|=0$. The system has to be cool
enough in order for quenching at work.
A numerical
simulation shows that domain structure is most prominent in the quenching
case \cite{ahw}.

\subsection{Numerical Simulations}
Numerical simulations have been done extensively in the literatures
\cite{dcc2}.
Usually the lattice spacing is chosen to be 0.5$\sim$1.0 fm, which cuts off
the momenta above 200$\sim$400 MeV, appropriate for studying soft pion modes.
In many cases, a boost invariance
in the longitudinal direction is assumed so that the longitudinal
expansion is automatically included.
\begin{figure}
\center
%\rule{2cm}{0.2mm}\hfill \rule{2cm}{0.2mm}
\vskip 8in
%\rule{2cm}{0.2mm}\hfill \rule{2cm}{0.2mm}
%\psfig{figure=radkaon.ps,height=1.5in}
\fcaption{This is what a DCC may look like! The top
one is the initial random configuration, the bottom one is the profile
of the field at $\tau = 5$ fm where two domains are clearly visible.
%Contour Plot of $\pi_2$ field in an event
%              at $\tau=\tau_0=1$ fm and $\tau = 5$ fm.
A quenching initial condition is used.}
\label{fig:radk}
\end{figure}
To monitor the domain formation, one has to compute the two point correlation
function. Figure 1 shows the evolution of an initially random noise with the
proper time $\tau$ using a quenching initial condition \cite{ahw}.
At $\tau =5$ fm, some domain structure clearly emerges. The typical domain
size can be as large as 2$\sim$3 fm in the transverse dimension and 1$\sim$2
unit in rapidity distribution.

\section{Signals of DCC}
\subsection{Anomalous Isospin Fluctuation}
The existence of the domain structure of the pion fields can be
most effectively observed in the phase space distribution of pion
multiplicity when the ``disorientation'' of individual domain is radiated away
in Goldstone modes (pions). The simplest way to quantize the semiclassical
field and to describe its decay is given in terms of a coherent state
\begin{eqnarray}
|\phi_{\rm cl}\rangle =N{\rm exp}\left\{ \int d^3k\phi_{\rm cl}(k)a^{\dag}
(k)\right\}|0\rangle \; .
\end{eqnarray}
There are some subtle issues on the conservation of quantum numbers in
a coherent state, while a squeezed state \cite{kogan}
seems to be more consistent in many ways.
Such coherent emission of semiclassical
pion field would lead to some unusual fluctuations in a given lego plot
acceptance sector $\Delta y\Delta \phi$. One may look for some unusually
rich or poor neutral pions compared to the charged pions in the given
sector. The Centauro and anti-Centauro events in the cosmic ray
experiments may well be a possible candidate for such signal.
If the interaction volume is small so that it contains only
one such domain as it may be the case in $pp$ collisions, it may be possible
to count event-to-event the ratio ($f$) of neutral to total pions yield .
If all orientations in a domain are equally possible, a simple calculation
shows \cite{dcc} that the probability  distribution ($n$) is
$\frac{1}{n}\frac{dn}{df}=\frac{1}{2\sqrt{f}}$,
very different from the normal binomial distribution.
This is what the T864 test/experiment at Tevatron led by Bjorken
and Taylor is currently looking for. To further reduce the background, one
may also apply a $p_t$ cut from above (say, 200 MeV) since after all the
domain size should be $\geq 1$ fm to be of any interest.

The theoretical simulation indicates that under suitable initial conditions,
preferably the quenching conditions,
the formation of sizable domains is quite possible. In a real experiment, the
signal will be diluted by the measure of the probability for such conditions
in all possible conditions and
the irreducible backgrounds. Therefore, the production of DCC may merely be
a rare event. Some accumulation of statistics may be needed.
The odds that the formation of domain favors a non-equilibrium
condition seems to suggest that a $pp$ or $p\bar{p}$
 collision has a better chance than
a heavy nucleus collision.
Currently, there has been no realistic simulation on
the cross section for the DCC production for the Tevatron Minimax project.

\subsection{Measuring Domain Size}
It is also very important to measure the size of domain which would reveal the
existence of a domain structure independently. The simplest way is to look at
the pion multiplicity as a function of $p_t$. For example, if the typical
domain size is $2$ fm, we should observe a rise of the multiplicity when
$p_t$ drops through $\sim$100 MeV.

\subsection{$\rho /\omega\rightarrow \ell^+\ell^-$ Decays}
It has been suggested \cite{hsw} that when the hadronic
resonances decay inside the
DCC, their electromagnetic decay modes show a clear
signal of the disorientation
of the background, serving as an alternative signature of DCC.
The idea is to use the fact that electromagnetic interactions break
the chiral symmetry explicitly (even in the chiral limit). The simplest way
to see this is that the $u$ and the $d$ quarks have different electric
charges, and as a result, $\pi^+$ and $\pi^0$ have different masses.

In the normal vacuum, the QCD vacuum aligns with the direction of
electromagnetic interactions. However, the DCC
does not in general align with the EM interactions.  One immediate
consequence of this misalignment is that the vector-meson
dominance no longer holds for the EM current in the familiar way.
In the disoriented domain, the meson state to which
the isovector current couples is no longer a single
meson state of mass $770$ MeV but
a linear combination of all possible charge states with mass $770$ MeV
and $1260$ MeV ($= m_{a_1}$).
If all
possible orientations of the chiral SU(2)$_L \times$SU(2)$_R$
are allowed with an equal probability,
on average, all six transitions, $\rho\rightarrow \gamma$
and $a_1 \rightarrow \gamma$, occur with an equal probability $1/6$.
This means that the $\rho$
peak in the $\ell^+\ell^-$ mass plot will be reduced by half
if all three charge states of $\rho$ are produced with an equal
rate, and a
new broad bump of width $\simeq 300$ MeV ($= \Gamma_{a_1}$)
may appear at 1260 MeV
\begin{eqnarray}
\Gamma\left( \rho(770 \mbox{MeV}) \rightarrow \ell^+\ell^-\right)_{\rm DCC}
& = &
\frac{1}{2} \Gamma\left( \rho^0 \rightarrow
\ell^+\ell^-\right)_{\rm normal}\; ,\\
\Gamma\left( a_1(1260 \mbox{MeV}) \rightarrow
\ell^+\ell^-\right)_{\rm DCC}
& = & \left( \frac{m_\rho}{m_{a_1}}\right) ^3
\Gamma\left( \rho(770 \mbox{MeV}) \rightarrow \ell^+\ell^-\right)_{\rm DCC}
\label{12}
\end{eqnarray}
This prediction is little affected by the current quark mass.
It is a consequence of the misalignment of the DCC with the
electroweak vacuum fixed by the Higgs field, so it would disappear
only at $F_\pi \rightarrow 0$ in which a distinction between the
$\rho$ and the $a_1$
ceases to be meaningful.

In contrast, the isoscalar
current is dominated by the $\omega$ meson ($780$ MeV) and the
$\phi$ meson ($1020$ MeV) in the normal vacuum, which are
the two singlets $({\bf 1}, {\bf 1})$ of SU(2)$_L \times$SU(2)$_R$.
Therefore, no matter how much the DCC vacuum is chirally disoriented,
the $\omega$ and $\phi$ mesons are not affected at all.  In addition,
because of its short lifetime, the $\rho$ decays mostly inside the DCC
domain while the $\omega$ lifetime is probably too long for it to decay
inside the DCC domain. Therefore there is a chance to test the
formation of the DCC by
carefully measuring the dilepton decays of the $\rho$ and the $\omega$
mesons.

\section{What We (Hope to) Learn}
We believe that if the idea of spontaneous symmetry-breaking is correct, the
vacuum is a quite complicated object. For a continuous symmetry, there
exist infinitely many degenerate (or almost degenerate) vacua. The fact that
we only live in one of them does not exclude us from being aware of others.
We know that there are Goldstone modes excited long the directions of
degeneracy. Pions are indeed very light.
Another possibility is to disturb the vacuum by a high energy
collision, where we are not using the high energy frontier to produce a
top quark pair, rather, converting the energy into large entropy, i.e.\
 producing many many low energy particles. These collision debris are
 distributed over a extended spatial region, interacting with the vacuum
using their long wavelength modes. Some of these modes may be unstable and tend
to grow exponentially, leading to the domains of different orientations in
this excited region.

This new phenomenon is called ``the vacuum engineering'' by
T.D.\ Lee \cite{lee}. A more concrete picture known as
``Baked Alaska'' has been suggested by Bjorken et al.\  who speculates
that a cool interior of large ``fireball'' may indeed
be produced in very high energy collisions.
In addition to the Minimax project, there have been some
experimental initiatives
searching for DCC at upcoming RHIC.  This will open up
not only an opportunity for studying the low energy
multiparticle physics in extremely high energy collisions, but also
a window for a direct test of the idea of spontaneous symmetry-breaking.
On this hopeful note, let me stop.

\section{Acknowledgements}
I wish to thank M.\ Asakawa, M.\ Suzuki and X.-N.\ Wang for their
fruitful collaborations. I am grateful to J.\ Bjorken and Y.\ Kluger
for many conversations on this subject. This work was
supported by the Director, Office of Energy Research,
Office of High Energy, Division of High Energy Physics
of the U.S.\
Department of Energy under contract DE-AC03-76SF00098
and the National Science and
Engineering Research Council of Canada.

\end{document}